\documentclass[twocolumn,slac_two]{revtex4}
\usepackage{graphicx}
\usepackage{fancyhdr}
\begin{document}

\title{Optical Outburst of the Gamma-Ray Blazar S4 0954+658 in March-April 2011}

%

\author{V.M.Larionov}
\affiliation{Astron. Inst. of St.Petersburg University, Russia}
\author{S.G. Jorstad, A.P. Marscher}
\affiliation{Boston University, USA}

\author{D.A. Morozova, I.S. Troitsky, D.A. Blinov, E.N. Kopatskaya, E.G. Larionova}
\affiliation{Astron. Inst. of St.Petersburg University, Russia}

\begin{abstract}
We present optical photopolarimetric observations of the BL Lac object S4\,0954+658 obtained with the 70-cm telescope in Crimea, 40-cm telescope in St.Petersburg, and 1.8-m Perkins telescope at Lowell Observatory (Flagstaff, Az). After a faint state with a brightness level R $\sim$17.6 mag registered in the first half of January 2011, the optical brightness of the source started to rise and reached $\sim$14.8 mag during the middle of March, showing flare-like behavior. The most spectacular case of intranight variability was observed during the night of 2011 March 9, when the blazar brightened by $\sim$0.7 mag within ~ 7 hours. During the rise of the flux the position angle of optical polarization rotated smoothly over more than 200 degrees. S4\,0954+658 is a gamma-ray blazar with gamma-ray flux of (5$\pm$3)x10$^{-10}$ phot\,cm$^{-2}$\,s$^{-1}$ according to the Fermi 11-month Catalog Extragalactic Sources. Our analysis of contemporaneous Fermi LAT data does not show any sign of increased gamma-ray activity above the detection threshold except for an elevated flux on 2011 March 5, JD\,2455626, coincident with the local optical maximum. 

\end{abstract}

\maketitle

\thispagestyle{fancy}


\section{INTRODUCTION}
S4\,0954+658 (z=0.367) is a well studied BL Lac object in the optical bands. Its optical variability has been studied by Wagner et al. (1993)~\cite{Wagner}, who found large amplitude variations (of the order of $\sim 100$\%) on time scales as short as 1 day. Raiteri et al. (1999)~\cite{Raiteri} presented a comprehensive study of the optical and radio variability of the source during 1994-1998. They detected large amplitude intranight variations. Studying the $B-R$    colour variations, they found that the mid and long-term variations in the source are not associated with spectral variations.
Gabuzda et al. (2000 and references therein)~\cite{Gabuzda} studied  the radio-morphology of S4\,0954+658  and showed that both the parsec and kiloparsec jet are bended. They also found substantial intranight polarization variability of the radio core. S4 0954+658 is a gamma-ray blazar with mean gamma-ray flux $(5\pm 3)\cdot 10^{-10} \mbox{ phot cm}^{-2} \mbox{s}^{-1}$ according to the Fermi 11-month Catalog of Extragalactic Sources, and most of time falls below the detection threshold.

\section{OPTICAL OBSERVATIONS}
Our photometric monitoring of S4\,0954+658 and other gamma-bright blazars is made within the framework of the WEBT/GASP project\footnote{\tt http://www.oato.inaf.it/blazars/webt}. We regularly   update optical light curves taken by St.Petersburg University team\footnote{\tt http://www.astro.spbu.ru/staff/vlar/OPTlist.html}.
We intensified our photometric and polarimetric observations of S4\,0954+658  following our detection of a new activity stage of this source (Larionov et al., 2011~\cite{Larionov}).

\begin{figure}
\includegraphics[width=80mm, clip]{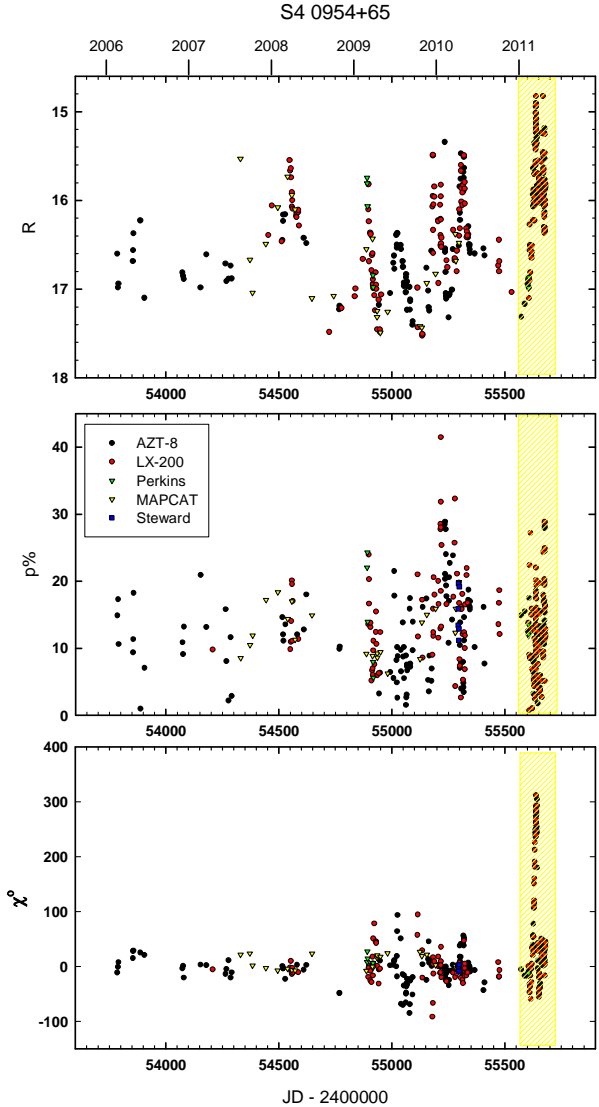}%
\caption{Optical (R band) brightness, fractional polarization and position angle of polarization of S4\, 0954+685 for 2006-2011.\label{fig1}}
\end{figure}

Figure~\ref{fig1} displays the whole set of optical photometric and polarimetric data obtained by our team during 2006-2011, supplemented by additional data from the MAPCAT project and Steward Observatory.

\begin{figure}
\includegraphics[width=80mm,clip]{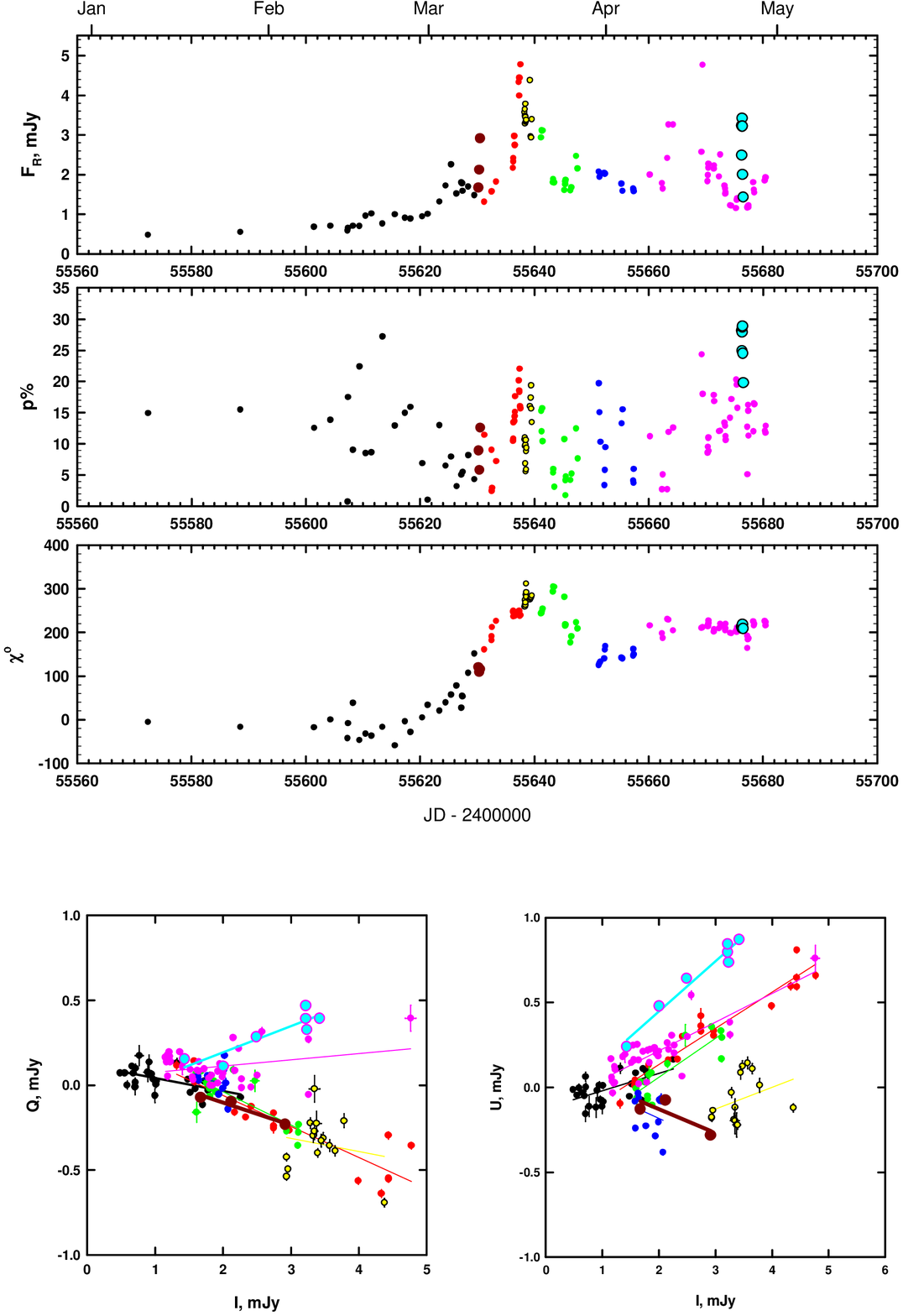}%
\caption{\textbf{Observed} flux density (corrected for galactic extinction), fractional polarization and position angle for January-May 2011. Larger symbols refer to the nights when violent intranight variability was observed.\label{fig2}}
\end{figure}

The blazar showed prominent activity during all the period covered with our data, with $R$ band amplitude exceeding $2^m$  and a record level of polarization  more than 40\% . Even on that background the outburst that started in early 2011 (marked yellow in all panels) is quite impressive. Its enlargement is shown in Fig.~\ref{fig2}.

Inspection by eye of Fig.\ref{fig1} shows that during most of the observational period the position angle of polarization (EVPA) was at a level of 0 degrees, close to the mean direction ($-10^\circ$) of the inner jet of S4\,0954+658, as observed by the VLBA. Unlike all the previous years, starting from the end of February 2011, we observed smooth rotation of  the EVPA (Fig.~\ref{fig2}, bottom panel), with amplitude of about 300 degrees. The maximum value of rotation was reached when the flux was at its peak. During two of the observational nights, March 9 and April 24, we observed violent  intranight variability, 0.7 mag within 7 hours and 1.0 mag within 5 hours, correspondingly, accompanied by synchronous changes in fractional polarization (marked with large symbols in Fig.~\ref{fig2}). As far as we know, these are the fastest flux and polarization changes recorded for this source.

\begin{figure}
\includegraphics[width=80mm,clip]{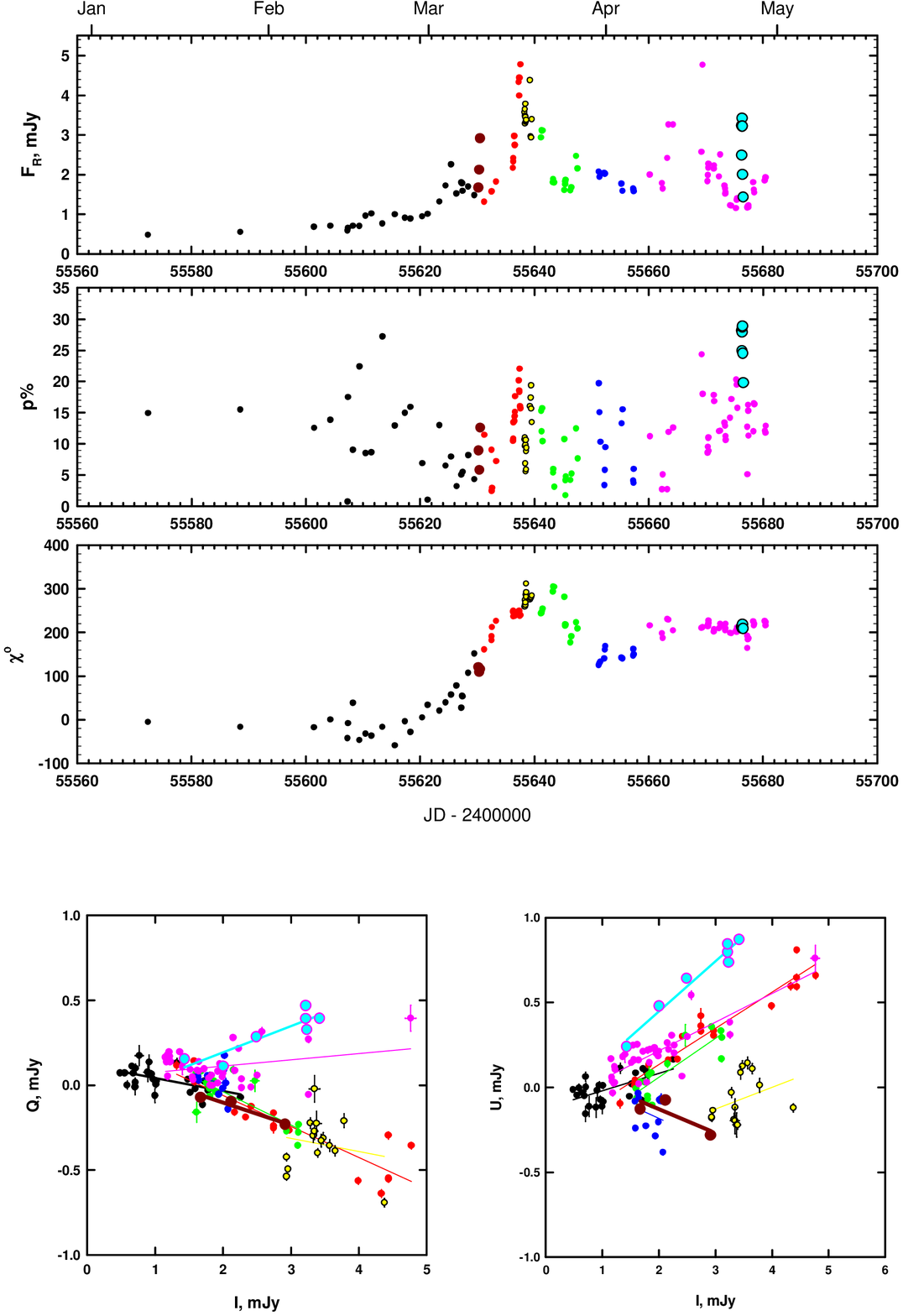}%
\caption{The absolute Stokes parameters variation during 2011 January-April. Different colors refer to different stages of evolution of the variable source.\label{fig3}}
\end{figure}

Following Hagen-Thorn and Marchenko  (1999)~\cite{Hagen}, we plotted ($Q$ vs $I$) and ($U$ vs $I$) Stokes polarization parameters (see Fig.~\ref{fig3}) and found that the whole data set can be splitted  into a succession of subsets, each with its own behavior in ($I, Q, U$) space. We mark these subsets with different colors and apply the same colors to the data plotted in Fig.~\ref{fig2}.

\begin{figure}
\includegraphics[width=80mm,clip]{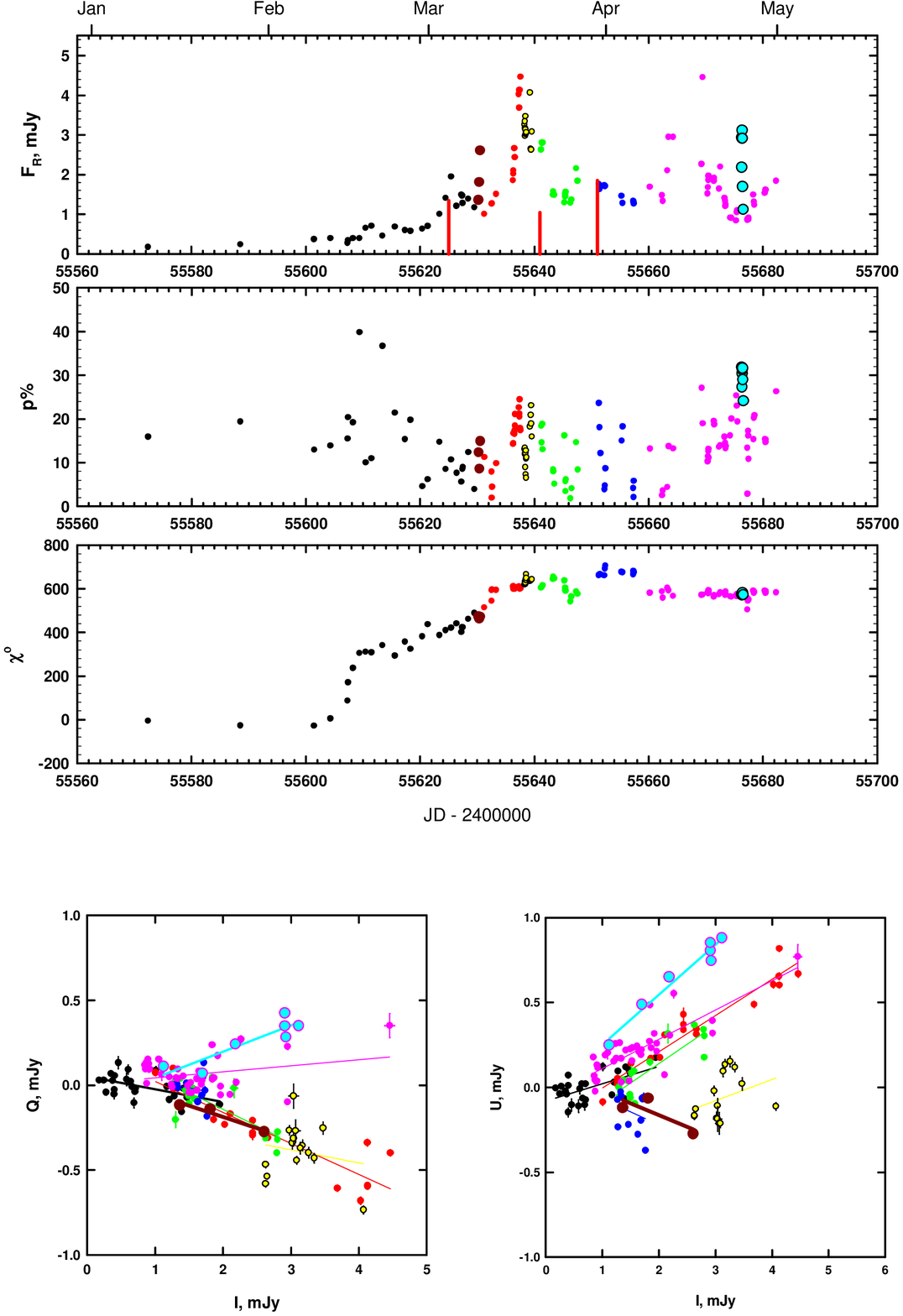}%
\caption{Same as Figure 2, but corrected for input of a constant component with R=18.7, p=15\% and $\chi= -6^\circ$. Red vertical bars in the upper panel mark positive Fermi LAT detections, the bars' height proportional to gamma-ray flux.\label{fig4}}
\end{figure}

We notice that the regression lines in Fig.~\ref{fig3} tend to converge to the points corresponding to the pre-outburst values of $I$, $Q$ and $U$. This means that the source that was probably seen before the outburst still was contributing the same amount of flux and polarization  during the outburst. Hence it is possible to subtract its emission to get the radiation parameters of the variable source(s). We estimated the constant source's parameters as R=17.8, p=15\% and $\chi= -6^\circ$. The resulting curves are given in Fig.~\ref{fig4}.

The position angle of polarization changed systematically. We see a sharp jump of EVPA at MJD~55609 (2011 Feb.16), followed by steady rotation $\sim$ 12.5 degrees/day that abruptly stopped at MJD~55632 (2011 Mar.11), 5 days before the optical maximum. After that only minor changes of EVPA are observed, despite continued strong variability of the flux density and fractional polarization. As for the previously mentioned jump of EVPA at MJD~55609, we believe that   this jump  may be an artifact caused by the appearance and increased contribution of the new variable polarized source with very different direction of electric vector. Note that this new-born source had very high (40\%) intrinsic polarization.

Most of our photometric observations were made in $BVRI$ bands. We looked for possible color variations, and found that 10-fold change of flux density was not accompanied by noticeable changes of colors.

\section{RADIO OBSERVATIONS}
The BL Lac object 0954+658 is monitored monthly by the BU group with the Very Long Baseline Array (VLBA) at 43 GHz within a sample of bright $\gamma$-ray blazars\footnote{\tt{http://www.bu.edu/blazars}}. 
Figure 5 shows the total and polarized intensity images of the object from 2010 September to 2011 March. The VLBA data were calibrated, imaged, and modeled in the same manner as discussed in Jorstad et al. (2005)~\cite{Jorstad}.

\begin{figure}
\includegraphics[width=80mm,clip]{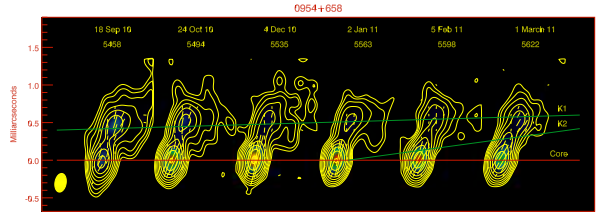}%
\caption{43 GHz total (contours) and polarized (color scale) intensity images of 0954+658, $S_{peak}=752$  mJy/beam, $S_{peakp}=101$ mJy/beam, beam=$0.24 \times 0.15$ $mas^2$ at $PA=-10^\circ$. Total intensity contours at 0.35, 0.7, 1.4....89.6\% of the peak. Sticks over the polarized intensity indicate the plane of polarization.\label{fig5}}
\end{figure}

\begin{figure}
\includegraphics[width=80mm,clip]{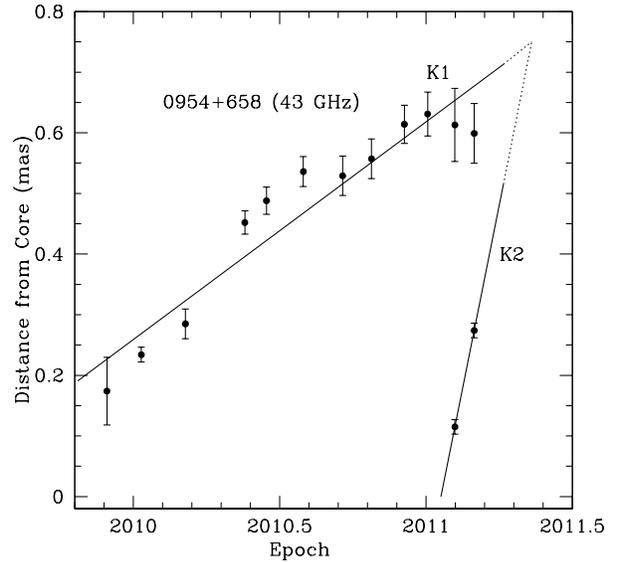}%
\caption{Separation of knots K1 and K2 from the core as function of time.\label{fig6}}
\end{figure}

Figure~\ref{fig5} shows 2 prominent polarized features, the core and knot K1. The position angle of polarization in the core is mostly along the jet (from -10 to -20 deg) while the EVPA in knot K1 is $\sim 0$ deg. At the last two epochs a new knot, K2, is imaged near the core. Figure~\ref{fig6} shows the motion of knots K1 and K2 with respect to the core (a presumably stationary feature) and approximate fit to the motion if the knots move ballistically.
K1 has a superluminal apparent speed $\beta_{app}=7.2\pm0.7 c$ ($H_0=70$, $\lambda_m=0.3$, $\lambda_\Omega=0.7$) and coincided with the core at epoch $2009.28\pm0.13$. The kinematics of K2 can be estimated only roughly since we have 2 epochs so far when the knot is visible, the data from which indicate that K2 is very highly relativistic, $\beta_{app}=19.1\pm 3.3$. An increase of total and polarized flux seen in K1 at the last epoch (March 1, 2011) can be attributed to K2 approaching K1 as well as a very fast optical outburst that just started to evolve at that epoch.

\section{GAMMA-RAY RESULTS}
\begin{figure}
\includegraphics[width=80mm,clip]{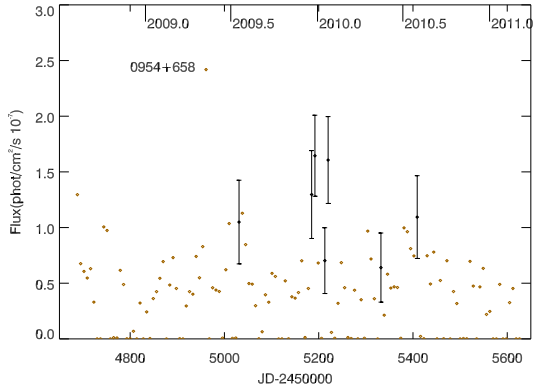}%
\caption{Fermi LAT gamma-ray light curve of S4\,0954+658 with 7-day binning of the data. Yellow dots mark upper limits.\label{fig7}}
\end{figure}

S4 0954+658 was only marginally detected by Fermi LAT during 2008-2010, as shown in Fig.~\ref{fig7}. A series of positive detections around 2010.0 may correspond to enhanced optical activity at that epoch (see Fig.~\ref{fig1}). We analyzed the LAT data for the period of 2011 January-April with 2-day binning, trying not to miss possible shortlived events, and found only 3 positive detections. All these cases of detection are centered on the highest optical level and are unlikely to be due to chance coincidences, though low signal-to-noise ratio precludes more definite conclusions.

\section{CONCLUSIONS}
S4\,0954+658 demonstrates very prominent activity starting from mid-February 2011. Our photometric and
polarimetric observations densely cover this period. We conclude that:

\begin{itemize}
\item{The overall behavior of the source can be explained as a superposition of radiation of a previously existing source with unchanged Stokes parameters and a new, strongly variable one, whose polarization vector rotated $\sim$ 12 degrees/day from the onset of the outburst until the moment of maximum flux and then leveled at $\sim 590^\circ$. Allowing for $k\cdot180^\circ$ ambiguity, this is equivalent to $50^\circ$, still quite different from the preoutburst direction ($-6^\circ$). This fast and monotonous rotation might be explained as the movement of the variable source in a helical magnetic field (Marscher et al. 2010).~\cite{Marscher}}
\item{High-amplitude intranight variations were superposed on both light and polarization curves. This may reflect fine structure of the magnetic field, causing temporary deflections.}
\item{The VLBA radio images show the appearance of a new highly relativistic component K2 with $\beta_{app} \sim 19$ c.}
\item{Fermi LAT data give indication of enhanced $\gamma$-ray activity close to the epoch of optical maximum.}
\end{itemize}

The period of activity of S4\,0954+658 still continues, and we hope that further optical and radio monitoring will allow us to construct a self-consistent picture of multi-wavelength behavior of this source.

\bigskip 
\begin{acknowledgments}
We thank Ivan Agudo and Paul Smith for the possibility to use their polarimetric data. The research at
St.Petersburg State U. is funded in part by RFBR grant 09-02-00092 and by the Ministry of Education and
Science of the Russian Federation (state contract \# P123). The research at BU is funded in part by NASA Fermi Guest Investigator grant NNX08AV65G and by NSF grant AST-0907893. The VLBA is an instrument of the
National Radio Astronomy Observatory, a facility of the National Science Foundation operated under cooperative agreement by Associated Universities, Inc.\end{acknowledgments}

\bigskip 

\end{document}